\documentclass[draftclsnofoot,onecolumn,journal,a4paper]{IEEEtran}%

\usepackage{cite}

\usepackage[pdftex]{graphicx}
\graphicspath{{pdf/}{jpeg/}}
\DeclareGraphicsExtensions{.pdf,.jpeg,.png}

\usepackage[cmex10]{amsmath}  % Use the [cmex10] option to ensure complicance
\usepackage{amssymb}
\usepackage{algorithmic}

\usepackage{array}

\ifCLASSOPTIONcompsoc
  \usepackage[caption=false,font=normalsize,labelfont=sf,textfont=sf]{subfig}
\else
  \usepackage[caption=false,font=footnotesize]{subfig}
\fi

\usepackage{stfloats}

\usepackage{flafter}

\usepackage{url}

\usepackage{cases}
\newcommand{\rb}[1] {\left(#1\right)}
\newcommand{\cb}[1] {\left\lbrace #1 \right\rbrace}
\newcommand{\sqb}[1] {\left[#1\right]}

\newcommand{\AgivenB}[2]{\left(#1\middle\vert#2\right)}
\newcommand{\ceil}[1]{\left\lceil#1\right\rceil}

\newtheorem{proposition}{Proposition}

\newtheorem{theorem}{Theorem}
\begin{document}
\title{Fundamental Latency Limits for D2D-Aided Content Delivery in Fog Wireless Networks} 

%%% Authors for arXiv
\author{
	Roy~Karasik,
	Osvaldo~Simeone,
	and~Shlomo~Shamai~(Shitz)
	\thanks{R. Karasik and S. Shamai are with the Faculty of Electrical Engineering, Technion - Israel Institute of Technology, Haifa, Israel (e-mails: royk@campus.technion.ac.il, sshlomo@ee.technion.ac.il). O. Simeone is with the Department of Informatics, King’s College of London, London, UK (e-mail: osvaldo.simeone@kcl.ac.uk).}% <-this % stops a space
	\thanks{This work has been supported by the European Research Council (ERC) under the European Union’s Horizon 2020 Research and Innovation Programme (Grant Agreement Nos. 694630 and 725731).}}

\maketitle

\begin{abstract}
  Device-to-Device (D2D) communication can support the operation of cellular systems by reducing the traffic in the network infrastructure. In this paper, the benefits of D2D communication are investigated in the context of a Fog-Radio Access Network (F-RAN) that leverages edge caching and fronthaul connectivity for the purpose of content delivery. Assuming offline caching, out-of-band D2D communication, and an F-RAN with two edge nodes and two user equipments, an information-theoretically optimal caching and delivery strategy is presented that minimizes the delivery time in the high signal-to-noise ratio regime. The delivery time accounts for the latency caused by fronthaul, downlink, and D2D transmissions. The proposed optimal strategy is based on a novel scheme for an X-channel with receiver cooperation that leverages tools from real interference alignment. Insights are provided on the regimes in which D2D communication is beneficial. 
\end{abstract}


\section{Introduction}
Device-to-Device (D2D) communication is a main enabler of novel applications such as mission critical communication, video sharing, and proximity-aware gaming and social networking. Furthermore, it can enhance conventional cellular services, including content delivery, by reducing the traffic at the cellular network infrastructure. D2D communication in cellular networks can be either out-of-band, whereby direct communication between the users takes place over frequency resources that are orthogonal with respect to the spectrum used for cellular transmission; or in-band, in which case the same frequency band is used for both D2D and cellular transmissions \cite{asadi2014survey}.

\begin{figure}[htbp]
	\centering
	\includegraphics[width=3.5in]{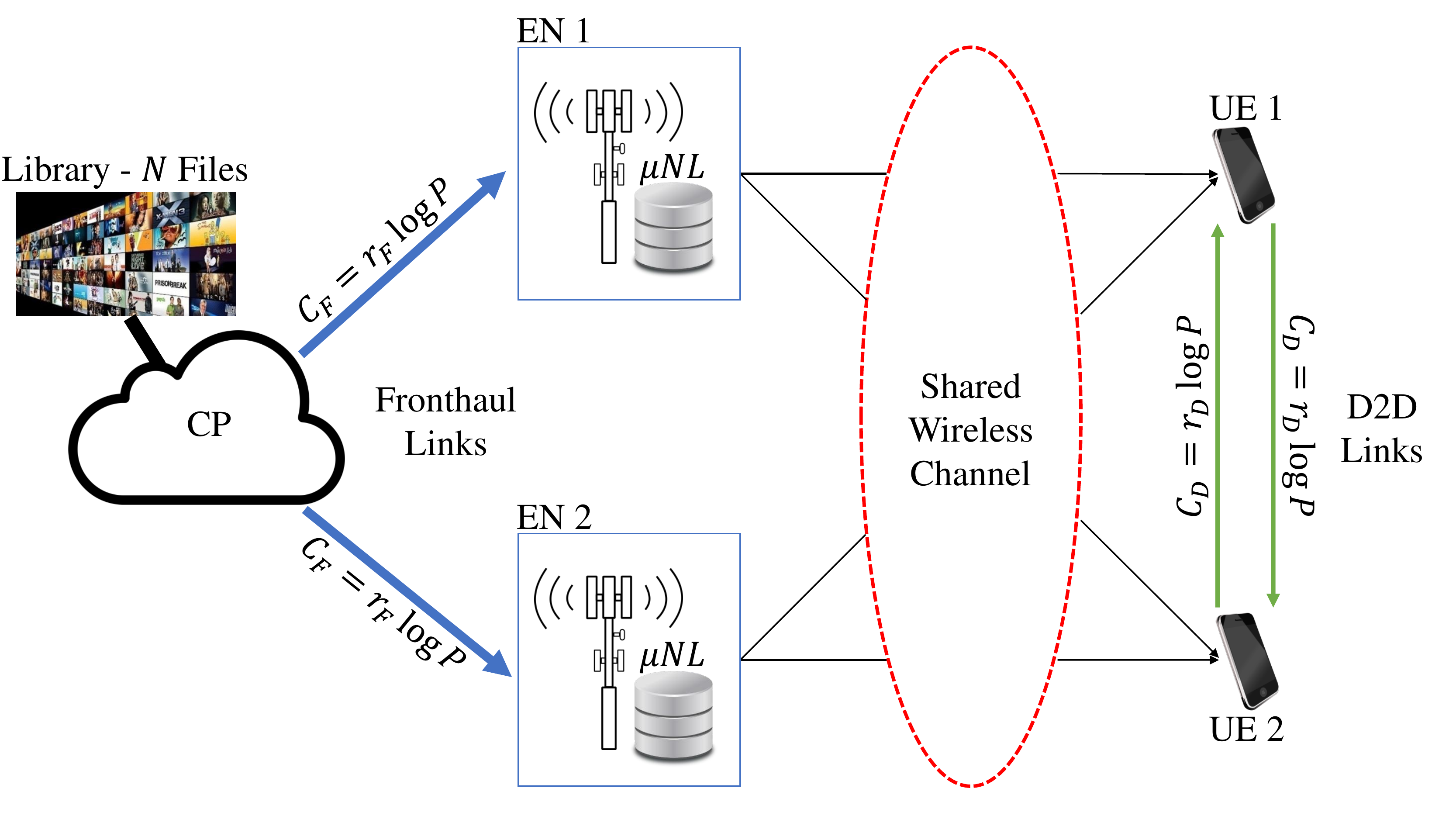}
	\caption{Illustration of the D2D-aided F-RAN model under study.}
	\label{fig_model}
\end{figure}

In this paper, we study the benefits of out-of-band D2D communications for the modern cellular architecture of a Fog-Radio Access Network (F-RAN) by focusing on content delivery \cite{hung2015architecture,tandon2016harnessing}. As illustrated in Fig \ref{fig_model}, in an F-RAN, content delivery leverages both edge caching and fronthaul connectivity to a Cloud Processor (CP). In this work, we characterize the potential latency reduction that can be achieved by utilizing D2D links in an F-RAN, while properly accounting for the latency overhead associated with D2D communications.

\textbf{Related Work:} The cache-aided interference channel was first studied in \cite{maddah2015cache}, where an upper bound on the minimum delivery latency in the high signal-to-noise ratio (SNR) regime was derived for a system with three users. A lower bound on the Normalized Delivery Time (NDT), which measures the high-SNR worst-case latency relative to an ideal system with unlimited caching capability, was presented in \cite{sengupta2016cache} for any number of Edge Nodes (ENs) and User Equipments (UEs), and it was shown to be tight for the setting of two ENs and two UEs. Lower and upper bounds for arbitrary numbers of ENs and UEs, where both ENs and UEs have caching capabilities, were presented in \cite{naderializadeh2017fundamental} under the constraint of linear precoders at the ENs. The NDT of a general F-RAN system with fronthaul links was studied in \cite{sengupta2016fog}, where the proposed schemes were shown to achieve the minimum NDT to within a factor of 2, and the minimum NDT was completely characterized for two ENs and two UEs, as well as for other special cases. In \cite{huang2009degrees}, it was shown that, for the interference channel with in-band cooperation, transmitter or receiver cooperation cannot increase the high-SNR performance in terms of sum Degrees of Freedom (DoF). The interference channel with out-of-band receiver cooperation was studied in \cite{wang2011interference}, where receiver cooperation was shown to increase the Generalized DoF metric. Importantly, reference \cite{wang2011interference} only imposes a rate constraint on the D2D links, hence not accounting for the latency overhead caused by D2D communications, which is of central interest in this work.

\textbf{Main Contributions:} In this paper, we study the D2D-aided F-RAN system with two ENs and two UEs in Fig. \ref{fig_model}, and put forth the following main contributions. First, in Sec. \ref{sec:X_channel}, we present a novel scheme that improves the NDT achievable on an X-channel with out-of-band D2D receiver cooperation. The proposed scheme enables interference cancellation at the receiver's side with minimal overhead on the D2D links. Second, in Sec. \ref{sec:minimum_NDT}, we characterize the minimum NDT of the D2D-aided F-RAN illustrated in Fig. \ref{fig_model}. The minimum NDT is used to identify the conditions under which D2D communication is beneficial, and to provide insights on the interplay between fronthaul and D2D resources.

\section{System Model}\label{sec:sys_model}
We consider the F-RAN system with Device-to-Device (D2D) links depicted in Fig. \ref{fig_model}, where two single-antenna User Equipments (UEs) are served by two single-antenna Edge Nodes (ENs) over a downlink wireless channel. The UEs are connected by two orthogonal out-of-band D2D links of capacity $C_D$ bits per symbol. The model generalizes the set-up studied in \cite{tandon2016cloud} by including D2D communications. Each EN is connected to a Cloud Processor (CP) by a fronthaul link of capacity $C_F$ bits per symbol. Throughout this paper, a symbol refers to a channel use of the downlink wireless channel. 

Let $\mathcal F$ denote a library of $N\geq 2$ files, $\mathcal F=\{f_1,\ldots,f_N\}$, each of size $L$ bits. The library is fixed for the considered time interval. 
The entire library is available at the CP, while the ENs can only store up to $\mu NL$ bits each, where $0\leq\mu\leq 1$ is the fractional cache size. During the placement phase, contents are proactively cached at the ENs, subject to the mentioned cache capacity constraints.

After the placement phase, the system enters the delivery phase, which is organized in Transmission Intervals (TIs). In every TI, each UE arbitrarily requests one of the $N$ files from the library. The UEs' requests in a given TI are denoted by the demand vector $\mathbf d\triangleq (d_1,d_2)\in[N]^2$, where for any positive integer $a$, we define the set $[a]\triangleq\{1,2,\ldots,a\}$. This vector is known at the beginning of a TI at the CP and ENs. The goal is to deliver the requested files to the UEs within the lowest possible delivery latency by leveraging fronthaul links, downlink channel and D2D links.

For a given TI, let $T_E$ denote the duration of the transmission on the wireless downlink channel. At time $t\in[T_E]$, each UE $k\in[2]$  receives a channel output given by 
\begin{IEEEeqnarray}{rCl}\label{eq:wireless_channel}
	y_k(t)&=&\sum_{m=1}^{2}h_{km}x_m(t)+z_k(t),
\end{IEEEeqnarray}
where $x_m(t)\in\mathbb C$ is the baseband symbol transmitted from EN $m\in[2]$ at time $t$, which is subject to the average power constraint $\mathbb E |x_m(t)|^2\leq P$ for some $P>0$; coefficient $h_{km}\in\mathbb C$ denotes the quasi-static flat-fading channel between EN $m$ to UE $k$, which is assumed to remain constant during each TI; and $z_k(t)$ is an additive white Gaussian noise, such that $z_k(t)\sim\mathcal C\mathcal N(0,1)$ is independent and identically distributed (i.i.d.) across time and UEs. The Channel State Information (CSI) $\mathbf{H}\triangleq\{h_{km}:k\in[2],m\in[2]\}$ is assumed to be drawn i.i.d. from a continuous distribution, and known to all nodes.

\subsection{Caching, Delivery and D2D Transmission}
The operation of the system is defined by the following policies that perform caching, as well as delivery via fronthaul, edge and D2D communication resources. 
\subsubsection{Caching Policy} During the placement phase, for EN $m$, $m\in[2]$, the caching policy is defined by functions $\pi^m_{c,n}(\cdot)$ that map each file $f_n$ to its cached content $s_{m,n}$ as
\begin{IEEEeqnarray}{rCl'l}\label{eq:caching_policy}
	s_{m,n}&\triangleq& \pi_{c,n}^m(f_n),&\forall n\in[N].
\end{IEEEeqnarray}
Note that, as per \eqref{eq:caching_policy}, we consider policies where only coding within each file is allowed, i.e., no inter-file coding is permitted. We have the cache capacity constraint $H(s_{m,n})\leq \mu L$. The overall cache content at EN $m$ is given by $s_m\triangleq(s_{m,1},s_{m,2}\ldots,s_{m,N})$. 
\subsubsection{Fronthaul Policy}In each TI of the delivery phase, for EN $m$, $m\in[2]$, the CP maps the library, $\mathcal F$, the demand vector $\mathbf{d}$ and CSI $\mathbf{H}$ to the fronthaul message
\begin{IEEEeqnarray}{rCl}
	\mathbf{u}_m=(u_m[1],u_m[2],\ldots,u_m[T_F])=\pi_f^m(\mathcal F,s_m,\mathbf{d},\mathbf{H}),
\end{IEEEeqnarray}
where $T_F$ is the duration of the fronthaul message. Note that the fronthaul message cannot exceed $T_FC_F$ bits, i.e., $H(\mathbf{u}_m)\leq T_FC_F$.
\subsubsection{Edge Transmission Policies}After fronthaul transmission, in each TI, the ENs transmit using a function $\pi_{e}^m(\cdot)$ that maps the local cache content, $s_m$, the received fronthaul message $\mathbf{u}_m$, the demand vector $\mathbf{d}$ and the global CSI $\mathbf{H}$, to the output codeword
\begin{IEEEeqnarray}{rClCl}
	\mathbf{x}_m&=&(x_m[1],x_m[2],\ldots,x_m[T_E])&=&\pi_{e}^m(s_m,\mathbf{u}_m,\mathbf{d},\mathbf{H}).
\end{IEEEeqnarray}
\subsubsection{D2D Interactive Communication Policies}After receiving the signals \eqref{eq:wireless_channel} over $T_E$ symbols, in any TI, the UEs use a D2D conferencing policy. For each UE $k\in[2]$, this is defined by the interactive functions $\pi^k_{\text{D2D},i}(\cdot)$ that map the received signal $\mathbf{y}_k\triangleq (y_k[1],\ldots,y_k[T_E])$, the global CSI and the previously received D2D message from UE $k'\neq k\in[2]$ to the D2D message
\begin{IEEEeqnarray}{l}\label{eq:d2d_policy}
	v_k[i]=\\
	\pi_{\text{D2D},i}^k(\mathbf{y}_k,\mathbf{H},v_{k'}[1],\ldots,v_{k'}[i-1],v_{k}[1],\ldots,v_{k}[i-1]),\IEEEnonumber
\end{IEEEeqnarray}
where $i=1,\ldots,T_D$, with $T_D$ being the duration of the D2D communication. The total size of each D2D message cannot exceed $T_DC_D$ bits. i.e., $H(v_k[1],\ldots,v_k[T_D])\leq T_DC_D$.
\subsubsection{Decoding Policy}After D2D communication, each UE $k\in[2]$ implements a decoding policy $\pi_d^k(\cdot)$ that maps the channel outputs, the D2D messages from UE $k'\neq k\in[2]$, the UE demand and the global CSI to an estimate of the requested file $f_{d_k}$ given as
\begin{IEEEeqnarray}{rCl}
	\hat{f}_{d_k}&=&\pi_d^k(\mathbf{y}_k,\mathbf{v}_{k'},d_k,\mathbf{H}).
\end{IEEEeqnarray}

The probability of error is defined as
\begin{IEEEeqnarray}{rCl}
	P_e&=&\max_{\mathbf{d}}\max_{k\in[2]}\Pr(\hat{f}_{d_k}\neq f_{d_k}),
\end{IEEEeqnarray}
which is the worst-case probability of decoding error measured over all possible demand vectors $\mathbf{d}$ and over all users $k\in[2]$. A sequence of policies, indexed by the file size $L$, is said to be feasible if, for almost all channel realization $\mathbf{H}$, we have $P_e\rightarrow0$ when $L\rightarrow\infty$.

\subsection{Performance Metric}
As discussed, in each TI, the CP first sends the fronthaul messages to the ENs for a total time of $T_F$ symbols; then, the ENs transmit on the wireless shared channel for a total time of $T_E$ symbols; and, finally, the UEs use the out-of-band D2D links for a total time of $T_D$ symbols. For any sequence of feasible policies, the delivery time per bit $\Delta(\mu,C_F,C_D,P)$ is hence defined as the limit
\begin{IEEEeqnarray}{rCl}\label{eq:del_time_per_bit}
	\Delta(\mu,C_F,C_D,P)&\triangleq&\limsup_{L\rightarrow\infty}\frac{\mathbb E(T_F+T_E+T_D)}{L}.
\end{IEEEeqnarray}
The notation emphasizes the dependence on the fractional cache size $\mu$, the fronthaul and D2D capacities $C_F$ and $C_D$, respectively, and the average power constraint $P$.

We adopt the Normalized Delivery time (NDT), introduced in \cite{sengupta2016fog}, as the performance metric of interest. To this end, we evaluate the performance in the high-SNR regime by parameterizing fronthaul and D2D capacities as $C_F=r_F\log(P)$ and $C_D=r_D\log(P)$. With this parametrization, the fronthaul rate $r_F\geq 0$ represents the ratio between the fronthaul capacity and the high-SNR capacity of each EN-to-UE wireless link in the absence of interference; and a similar interpretation holds for the D2D rate $r_D\geq0$.
  
For any given tuple $(\mu,r_F,r_D)$, the NDT of a sequence of achievable policies is defined as
\begin{IEEEeqnarray}{rCl}\label{eq:NDT_def}
	\delta(\mu,r_F,r_D)&\triangleq&\lim_{P\rightarrow\infty}\dfrac{\Delta(\mu,r_F\log(P),r_D\log(P),P)}{1/\log(P)}.
\end{IEEEeqnarray}
The factor $1/\log(P)$, used for normalizing the delivery time in \eqref{eq:NDT_def}, represents the minimal time to deliver one bit over an EN-to-UE wireless link in the high-SNR regime and in the absence of interference. The minimum NDT is finally defined as the minimum over all achievable policies
\begin{IEEEeqnarray}{l}\label{eq:minimum_ndt_def}
	\delta^*(\mu,r_F,r_D)\triangleq\IEEEnonumber\\
	\qquad\inf\{\delta(\mu,r_F,r_D):\delta(\mu,r_F,r_D)\text{ is achievable}\}.
\end{IEEEeqnarray} 
By construction, we have the lower bound $\delta^*(\mu,r_F,r_D)\geq 1$. Furthermore, the minimum NDT can be proved by means of time- and memory-sharing arguments to be convex in $\mu$ for any fixed values of $r_F$ and $r_D$ \cite[Remark 1]{tandon2016cloud}.

\section{The Two-User X-Channel with Receiver Cooperation}\label{sec:X_channel}
\begin{figure}[htbp]
	\centering
	\includegraphics[width=2.5in]{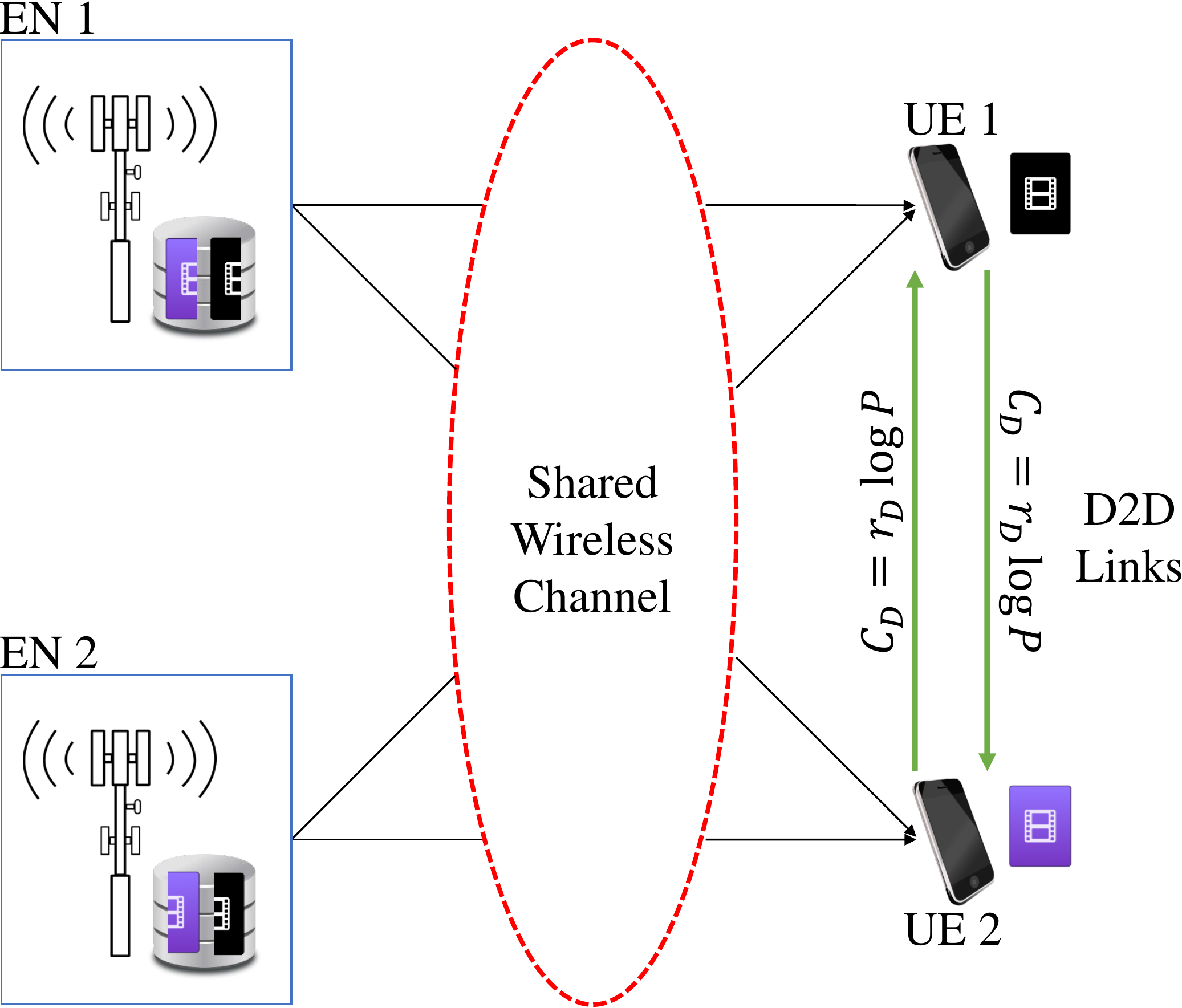}
	\caption{X-channel with receiver cooperation studied in Sec. \ref{sec:X_channel}, which represents an F-RAN system with no fronthaul, i.e., with $r_F=0$, and fractional cache size $\mu=1/2$.}
	\label{fig:X_chan}
\end{figure}

In this section, we present a result of independent interest that will be used in Sec. \ref{sec:minimum_NDT} to derive the minimum NDT \eqref{eq:minimum_ndt_def}. Specifically, we develop a new delivery scheme for the special case in which no fronthaul communication is enabled, i.e., $r_F=0$, and the fractional cache size is $\mu=1/2$. In this regime, each EN can only store half of each file in the library. Under the mentioned caching strategy, in the worst-case scenario in which the UEs request different files, the set-up is equivalent to a two-user Gaussian X-channel with receiver cooperation. In this channel, as illustrated in Fig. \ref{fig:X_chan}, each UE needs to download half of the requested file from one EN and the other half from the second EN. 
The proposed scheme achieves the NDT detailed in the following Proposition. 
\begin{proposition}\label{prop:x_ch_ub}
	For $\mu=1/2$, $r_F=0$ and $r_D\geq 0$, the minimum NDT is upper bounded as
	\begin{IEEEeqnarray}{rCl}\label{eq:x_ch_ub}
		\delta^*\rb{\mu=\frac{1}{2},r_F=0,r_D}\leq \delta_X\triangleq 1+\frac{1}{2r_D}.
	\end{IEEEeqnarray}
\end{proposition}

Proposition \ref{prop:x_ch_ub} is proved in the next two subsections by first proposing a novel scheme for the deterministic X-channel, and then adapting it for the Gaussian counterpart. The scheme is based on layered transmission and successive interference cancellation at the receivers.

As compared to existing schemes that are applicable for $\mu=1/2$ and $r_F=0$, real interference alignment \cite{motahari2014real} achieves an NDT of $3/2$ without using the D2D links \cite{tandon2016cloud}. Therefore, the proposed D2D-based scheme of Proposition \ref{prop:x_ch_ub} is useful only when the D2D capacity is sufficiently large, i.e., when $r_D>1$. Furthermore, the scheme in \cite{wang2011interference} has an NDT lower bounded by 2, since the latency due to D2D communications equals the transmission time on the downlink channel. Hence, the scheme is not advantageous in terms of NDT. Finally, as an alternative policy, one could have each UE compress and forward the received signal to the other UE, allowing each UE to carry out Zero Forcing (ZF) linear equalization. By quantizing with a rate equal to $\log P$, one can ensure that the SNR scales linearly with $P$, and that the approach achieves an NDT equal to $1+1/r_D>\delta_X$ (see \cite{sengupta2016fog}\cite{tandon2016cloud} for similar arguments).

\subsection{The Deterministic Approach}\label{subsec:deterministic_approach}
\begin{figure}[htbp]
	\centering
	\includegraphics[width=2in]{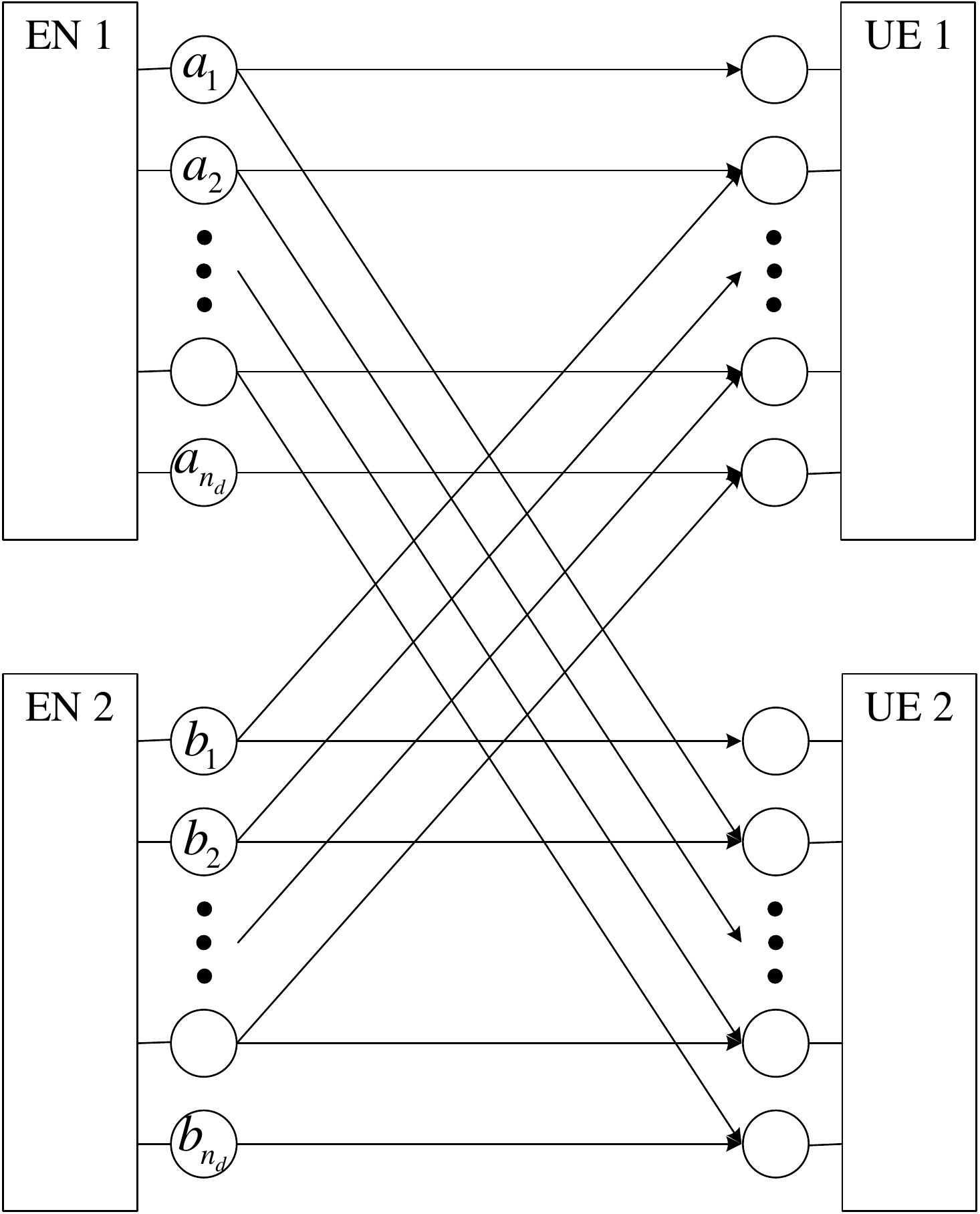}
	\caption{Deterministic X-channel with $n_d$ direct signal levels and $n_c=n_d-1$ cross signal levels considered in Sec. \ref{subsec:deterministic_approach}.}
	\label{fig:det_chan}
\end{figure}

We start by considering a deterministic approximation of the X-channel in order to facilitate the explanation of the main ideas behind the proposed scheme.
We recall that, according to \cite{huang2012interference}, in high SNR, the channel \eqref{eq:wireless_channel} is approximated by the deterministic model 
\begin{IEEEeqnarray}{rCl}\label{eq:deterministic_channel}
	y_1(t)&=&x_1(t)+S^{n_d-n_c}x_2(t)\nonumber\\
	y_2(t)&=&S^{n_d-n_c}x_1(t)+x_2(t),
\end{IEEEeqnarray}
where summations and multiplications are over the binary field $\mathbb F_2$; $n_d$ and $n_c$ represent the number of direct and cross signal levels, respectively, with $n_d>n_c$; $x_i(t)$ and $y_i(t)$ $\in\mathbb F_2^{n_d}$ for $i\in[2]$ are the binary vectors representing the inputs and outputs of the deterministic channel, respectively; and $S$ is the $n_d\times n_d$ shift matrix with all zeros except in the first lower diagonal, which contains all ones.
This channel is illustrated in Fig. \ref{fig:det_chan}.
The number of levels is selected as $n_d=\ceil{\log P}$, while $n_c$ will be taken to satisfy the limit $n_c/n_d\rightarrow 1$ when $n_d\rightarrow\infty$ in order to approximate the high-SNR behavior of the assumed channel model \eqref{eq:wireless_channel}, as explained in \cite[Appendix B]{huang2012interference}. Following this model, we set the D2D link capacity $C_D=r_D\log P$ to equal $r_Dn_d$ signal levels between the UEs.

Consider, without loss of generality, the case where $n_c=n_d-1$ and $n_d$ is odd, as illustrated in Fig. \ref{fig:det_chan}.
EN 1 and EN 2 at each time $t$ transmit independent bits $x_1=[a_1,\ldots,a_{n_d}]^T$ and $x_2=[b_1,\ldots,b_{n_d}]^T$ on the $n_d$ levels, where we have dropped the dependence on $t$. By \eqref{eq:deterministic_channel}, the received signals at the UEs are $y_1=[a_1,a_2\oplus b_1,\ldots,a_{n_d}\oplus b_{n_d-1}]^T$ and  $y_2=[b_1,b_2\oplus a_1,\ldots,b_{n_d}\oplus a_{n_d-1}]^T$.
UE 2 uses its D2D link to convey the bits received on the even-numbered levels
\begin{IEEEeqnarray}{rCl}\label{eq:det_d2d_msg}
	v_2&=&\cb{b_2\oplus a_1,b_4\oplus a_3,\ldots,b_{n_d-1}\oplus a_{n_d-2}}
\end{IEEEeqnarray}
to UE 1, which consists of $(n_d-1)/2$ bits. UE 1 is thus able to decode the bits $\{a_1,b_2,a_3,b_4,a_5,\ldots,b_{n_d-1},a_{n_d}\}$ from $\{y_1,v_2\}$ by means of successive interference cancellation. To this end, it starts by decoding $a_1$ from $y_{1,1}=a_1$; then, it uses $a_1$ together with $b_2\oplus a_1$ in \eqref{eq:det_d2d_msg} to decode $b_2$; next, it uses $b_2$ and $y_{1,3}=a_3\oplus b_2$ to decode $a_3$; and so on, until all the desired bits are decoded. Similarly, UE 2 is able to decode bits $\{b_1,a_2,b_3,a_4,b_5,\ldots,a_{n_d-1},b_{n_d}\}$ from $y_2$ and $v_1=\{a_2\oplus b_1,a_4\oplus b_3,\ldots,a_{n_d-1}\oplus b_{n_d-2}\}$. 
%Note that the scheme does not require UEs' interactions, but only two simultaneous one-shot D2D communications.

The number of channel uses required on the downlink channel to satisfy the UEs' demands is $L/(n_d-1)$. For each channel use, each UE has to convey $(n_d-1)/2$ bits using a D2D link of capacity $r_Dn_d$. Therefore, the resulting NDT \eqref{eq:NDT_def}, if we let the number $n_d$ of levels be arbitrary, is
\begin{IEEEeqnarray}{rCl}
	\lim_{n_d\rightarrow\infty}\frac{n_d}{n_d-1}\rb{1+\frac{(n_d-1)/2}{r_Dn_d}}=\delta_X.
\end{IEEEeqnarray}
Next, we show how to achieve the same NDT for the original model \eqref{eq:wireless_channel}.

\subsection{Real Interference Alignment with Receiver Cooperation}\label{sec:real_IA}
In order to convert the proposed scheme from the deterministic model to the X-channel \eqref{eq:wireless_channel}, we follow the \textit{real interference alignment} approach of \cite{motahari2014real}. Accordingly, in a manner similar to the deterministic model, each transmitter uses $n_d$ signal layers, where $n_d$ is odd. The signal transmitted by the ENs at each symbol can be written as
\begin{IEEEeqnarray}{c}\label{eq:layers}
	x_1=\sum_{i=1}^{n_d} g_{1,i}a_i\text{ and }x_2=\sum_{i=1}^{n_d} g_{2,i}b_i,
\end{IEEEeqnarray}
where $\{g_{m,i}\}$, with $m\in[2]$ and $i\in[n_d]$, are precoder gains, and the values $a_i$ and $b_i$ are chosen from a discrete constellation, so that we have $a_i,b_i\in A\mathbb Z_Q\triangleq \{0,A,2A,\ldots,A(Q-1)\}$. 
Each layer $i$ is coded using random coding with rate $R$ bits per symbol. Appendix \ref{sec:appendixA} shows that, by choosing parameters $\{g_{m,i}\}$, $A$, $Q$ and $R$ properly, UE 1 can decode the symbols $\{a_1,a_2+b_1,\ldots,a_{n_d}+b_{n_d-1},b_{n_d}\}$, while UE 2 decodes  $\{b_1,b_2+a_1,\ldots,b_{n_d}+a_{n_d-1},a_{n_d}\}$. The UEs now exchange the even-numbered layers as in the deterministic model, so that UE 1 transmits the message $v_1=\{a_2+b_1,a_4+b_3,\ldots,a_{n_d-1}+b_{n_d-2}\}$ to UE 2, while UE 2 transmits $v_2=\{b_2+a_1,b_4+a_3,\ldots,b_{n_d-1}+a_{n_d-2}\}$ to UE 1. As a result, UE 1 can decode $\{a_1,b_2,a_3,b_4,\ldots,a_{n_d},b_{n_d}\}$ while UE 2 decodes $\{b_1,a_2,b_3,a_4,\ldots,b_{n_d},a_{n_d}\}$.

As also shown in Appendix \ref{sec:appendixA}, for high SNR, the rate can be selected as $R\approx\log Q\approx \log P/(n_d+1)$. Since each UE decodes $(n_d+1)/2$ layers from one EN and $(n_d-1)/2$ from the other, the number of channel uses required to satisfy the UEs' demands on the edge channel is 
\begin{IEEEeqnarray}{rCl}
	T_E=\frac{L/2}{\frac{\log P}{n_d+1}\cdot\frac{n_d-1}{2}}=\frac{L}{n_d-1}\cdot\frac{n_d+1}{\log P}.
\end{IEEEeqnarray}
For each channel use, the message $v_k$, $k\in[2]$, conveyed over the D2D link, comprises $(n_d-1)/2$ elements, each taking one of $2Q$ values. Therefore, for high SNR, the delay due to D2D transmissions is given as
\begin{IEEEeqnarray}{rCl}
	T_D=T_E\cdot\frac{\log(2Q)\cdot(n_d-1)/2}{r_D\log P}\approx T_E\cdot \frac{n_d-1}{n_d+1}\cdot \frac{1}{2r_D},\quad
\end{IEEEeqnarray}
and the resulting NDT \eqref{eq:NDT_def} is
\begin{IEEEeqnarray}{rCl}\label{eq:fading_X_NDT}
	\delta_{n_d} \triangleq \frac{n_d+1}{n_d-1}\cdot\rb{1+\frac{(n_d-1)/2}{r_D(n_d+1)}}.
\end{IEEEeqnarray}
Similar to the deterministic channel, by increasing the number of layers $n_d$, we have the limit $\lim_{n_d\rightarrow\infty}\delta_{n_d}=\delta_X$.

\section{Minimum NDT}\label{sec:minimum_NDT}
In this section, we derive the minimum NDT by presenting a novel achievable scheme and an information-theoretic lower bound. The achievable scheme leverages the D2D cooperative strategy introduced above along with the scheme in \cite{tandon2016cloud}, which is optimal in the absence of D2D links.
\begin{theorem}\label{th:minimum_NDT}
	The minimum NDT for the $2\times 2$ F-RAN system with number of files $N\geq 2$, fractional cache size $\mu\geq0$, fronthaul rate $r_F\geq 0$ and D2D rate $r_D\geq 0$ is given as
	\begin{IEEEeqnarray}{ll}\label{eq:minimum_NDT}
		&\delta^*\rb{\mu,r_F,r_D}=\\
		&\left\lbrace\begin{array}{ll}
			\max\cb{1+\mu+\frac{1-2\mu}{r_F},2-\mu}&\text{for }0\leq r_F,r_D\leq 1\IEEEyesnumber\IEEEyessubnumber\\
			1+\frac{1-\mu}{r_F}&\text{for } r_F\geq\max\cb{1,r_D}\\
			\max\cb{1+\frac{\mu}{r_D}+\frac{1-2\mu}{r_F},1+\frac{1-\mu}{r_D}}&\text{for } r_D>\max\cb{1,r_F}.
		\end{array}\right.\IEEEnonumber
	\end{IEEEeqnarray}
\end{theorem}

\begin{figure}[htbp]
	\centering
	\includegraphics[width=3in]{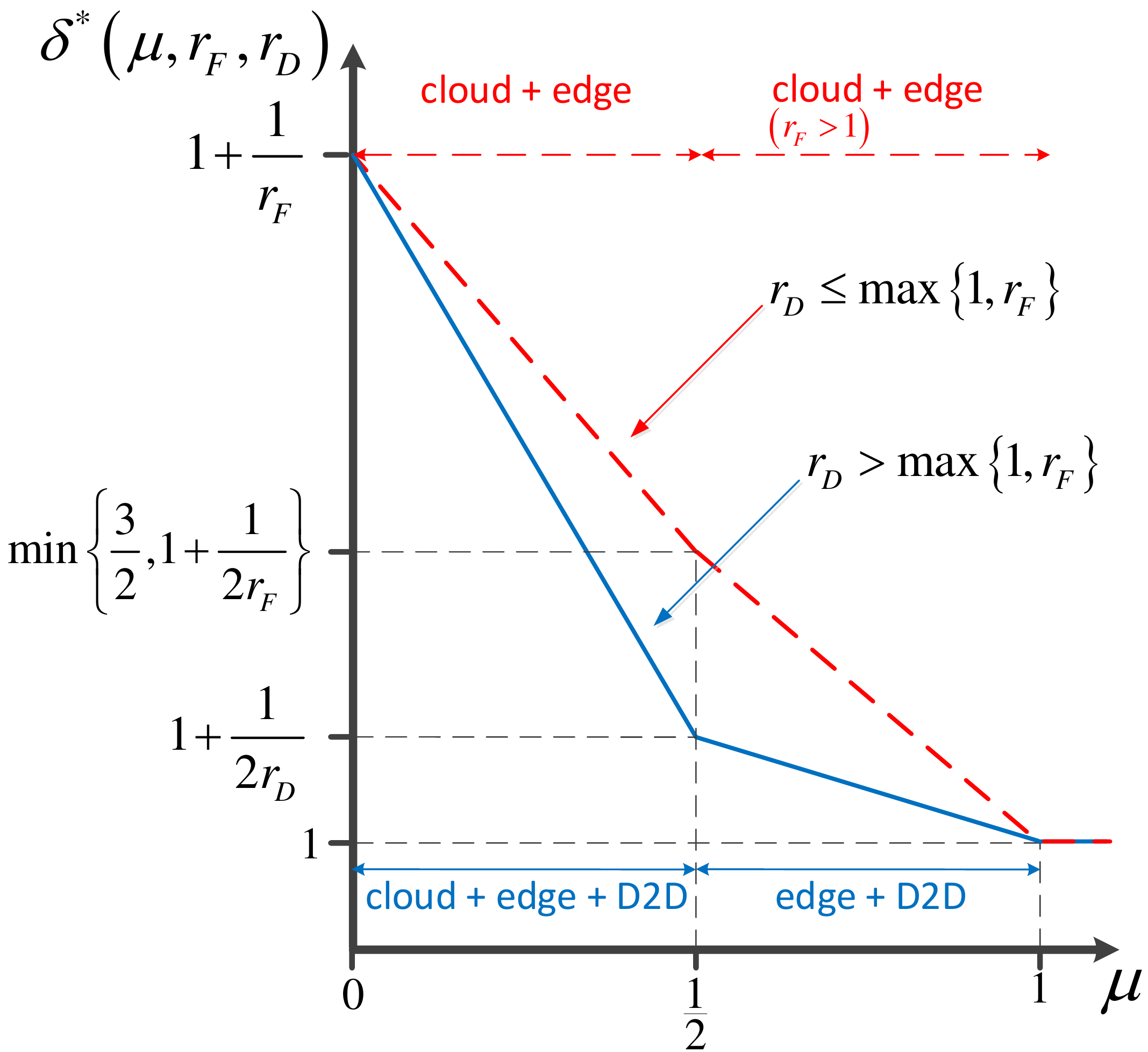}
	\caption{Minimum NDT for the $2\times 2$ F-RAN with D2D links as a function of $\mu$: when $r_D\leq\max\cb{1,r_F}$, D2D communication cannot reduce the delivery latency, while a reduction of the NDT is obtained when $r_D>\max\cb{1,r_F}$.}
	\label{fig:minimum_NDT}
\end{figure}

Before sketching the proof, we use the result in Theorem \ref{th:minimum_NDT} in order to draw conclusions on the role of D2D cooperation in improving the delivery latency. We start by observing that, for $r_D\leq\max\{1,r_F\}$, the minimum NDT \eqref{eq:minimum_NDT} is identical to the minimum NDT without D2D links derived in \cite[Theorem 1]{tandon2016cloud}. Therefore, D2D communication provides a latency reduction only when we have $r_D>\max\{1,r_F\}$. This is illustrated in Fig. \ref{fig:minimum_NDT}, where we plot the minimum NDT \eqref{eq:minimum_NDT} as a function of the fractional cache size $\mu$ for fixed fronthaul rate $r_F$ and D2D rate $r_D$. For any $r_D\leq\max\{1,r_F\}$, the minimum NDT is not affected by the value of $r_D$, whereas a larger $r_D$ yields a reduced minimum NDT.

The minimum useful value $\max\{1,r_F\}$ for the D2D rate $r_D$ increases with fronthaul rate $r_F$. This demonstrates that there exists a trade-off between fronthaul and D2D resources for the purpose of interference management, although their role is not symmetric. The use of fronthaul links is in fact necessary to obtain a finite NDT when the library is not fully available at the ENs, i.e., when $\mu<1/2$. D2D links can instead only reduce the NDT in regimes where fronthaul and edge resources would already be sufficient for content delivery with a finite NDT. In particular, as summarized in Fig. \ref{fig:minimum_NDT}, when $r_D>\max\{1,r_F\}$, D2D communication reduces the minimum NDT for all values $0<\mu<1$. Furthermore, when $\mu>1/2$, irrespective of the value of $r_F$, the minimum NDT is achieved by leveraging only edge caching and D2D links, without having to rely on fronthaul resources, thus reducing the traffic at the network infrastructure. This is in contrast to the case $r_D\leq\max\{1,r_F\}$, where, by \cite{tandon2016cloud}, fronthaul transmission is needed to obtain the minimum NDT unless $r_F\leq 1$.

\textit{Achievability:} The strategy that achieves \eqref{eq:minimum_NDT} is based on time- and memory-sharing \cite[Remark 1]{tandon2016cloud} between the policies for the corner points $\mu=0,1/2$ and $1$. For $\mu=1$, we apply cache-aided cooperative ZF at the ENs by leveraging the fact that the ENs can both store the entire library of files. This achieves the NDT $\delta^*(\mu=1,r_F,r_D)=1$ \cite[Sec. IV.A]{tandon2016cloud}. For $\mu=0$, we apply the cloud-aided soft-transfer scheme of \cite{sengupta2016fog} and \cite{tandon2016cloud}, which uses fronthaul links to convey quantized ZF-precoded signals, achieving the NDT $1+1/r_F$. Finally, for $\mu=1/2$, we use one of the following three schemes: \textit{(i)} EN coordination via interference alignment, which results in an NDT of 3/2 without using either fronthaul or D2D links \cite[Sec. IV.A]{tandon2016cloud}; \textit{(ii)} time- and memory-sharing between cloud-aided soft transfer and cache-aided cooperative ZF, which leverages fronthaul and cache resources, and results in an NDT of $1+1/(2r_F)$ \cite[Theorem 1]{tandon2016cloud}; \textit{(iii)} the proposed D2D-based delivery scheme, which results in an NDT of $\delta_X$ by leveraging edge and D2D resources. 

\textit{Converse}: The proof of the lower bound can be found in Appendix \ref{sec:appendix_lb}. The proof leverages the approach of \cite{tandon2016cloud}, which is based on a variation of cut-set arguments. Accordingly, subsets of information resources are identified, from which, in the high-SNR regime, the requested files can be reliably decoded when a feasible policy is implemented. In particular, the first subset, $\{s_1,s_2,\mathbf{u}_1,\mathbf{u}_2\}$, yields a lower bound on $T_F$ as a function of $\mu$ and $r_F$; the seconds subset, $\{\mathbf{y}_1,\mathbf{y}_2\}$, yields a lower bound on $T_E$; and the third subset, $\{s_1,\mathbf{u}_1,\mathbf{y}_1,\mathbf{v}_2\}$, where $\mathbf{v}_2\triangleq(v_2[1],\ldots,v_2[T_D])$, yields a lower bound on a linear combination of $(T_F,T_E,T_D)$ as a function of $\mu$, $r_F$ and $r_D$. Note that, only the latter bound differs with respect to \cite{tandon2016cloud}. These bounds are then linearly combined according to the values of the fronthaul rate $r_F$ and D2D rate $r_D$ to show that \eqref{eq:minimum_NDT} is a lower bound on the minimum NDT.

\textit{Remark:} Although the definition of D2D conferencing policy \eqref{eq:d2d_policy} allows for \textit{interactive communication} \cite{willems1983discrete}, the optimal scheme described above uses \textit{simultaneous monologues}, whereby the D2D messages of the UEs are based only on the CSI and the respective own received signal.

\section{Conclusions}
In this work, fundamental insights were provided on the benefits of D2D communication for content delivery in an F-RAN. 
Considering the Normalized Delivery Time (NDT) metric, an optimal strategy for utilizing the fronthaul and D2D links, as well as the downlink wireless channel, was presented. This strategy is based on a novel scheme for the X-channel with receiver cooperation. It was demonstrated that, for sufficiently large D2D and cache capacities, D2D communication can reduce the traffic on the fronthaul links, and hence help reducing the load on the network infrastructure. 
Among possible extensions of this work, we mention the generalization of the results to more than two users and ENs; the study of pipelined transmission, where fronthaul or D2D links can be used simultaneously with the downlink channel \cite{sengupta2016fog}; the evaluation of the impact of imperfect CSI; and the characterization of the optimal D2D strategy under linear precoding and hard-transfer constraints \cite{zhang2017fundamental}.

\appendices

\section{Proof of Proposition \ref{prop:x_ch_ub}}\label{sec:appendixA}
In the layered transmitted signals \eqref{eq:layers}, the precoder gains $\{g_{m,i}\}$ are chosen as
\begin{IEEEeqnarray}{rCl}
	g_{m,i}&=&\left\lbrace\begin{array}{ll}
		\rb{h_{11}h_{22}}^{\frac{n_d-i}{2}}\cdot\rb{h_{12}h_{21}}^{\frac{i-1}{2}}&\text{for }i\text{ odd}\\
		\rb{h_{11}h_{22}}^{\frac{n_d-i-1}{2}}\cdot\rb{h_{12}h_{21}}^\frac{i-2}{2}h_{m',m'}h_{m,m'}&\text{for }i\text{ even},
	\end{array}\right.\IEEEnonumber\\
\end{IEEEeqnarray}
where $m,m'\in[2]$ and $m'\neq m$. This choice results in layers $a_i$ and $b_{i-1}$ being summed at UE 1, while layers $b_i$ and $a_{i-1}$ are summed at UE 2, $i\in\{2,3,\ldots,n_d\}$. This is in the sense that, given the equalities $h_{11}g_{1,i}=h_{12}g_{2,i-1}$ and $h_{22}g_{2,i}=h_{21}g_{1,i-1}$, the received signals $y_1$ and $y_2$ can be written as
\begin{IEEEeqnarray}{rCl}\label{eq:x_ch_rec_ue1}
	y_1&=&h_{11}g_{1,1}a_1+\sum_{i=2}^{n_d}h_{11}g_{1,i}\rb{a_i+b_{i-1}}\IEEEnonumber\\
	&&+h_{12}g_{2,n_d}b_{n_d}+z_1,
\end{IEEEeqnarray}
and
\begin{IEEEeqnarray}{rCl}\label{eq:x_ch_rec_ue2}
	y_2&=&h_{22}g_{2,1}b_1+\sum_{i=2}^{n_d}h_{22}g_{2,i}\rb{b_i+a_{i-1}}\IEEEnonumber\\
	&&+h_{21}g_{1,n_d}a_{n_d}+z_2.
\end{IEEEeqnarray} 
Since symbols $a_i$ and $b_i$ are selected from $A\mathbb Z_Q$, the noiseless received signal $y_k-z_k$ takes values from a discrete set as well. Let $d_{\text{min},k}$ denote the minimum distance between the element of this set, and define $d_\text{min}\triangleq\min\cb{d_{\text{min},1},d_{\text{min},2}}$.
Almost surely, the effective channels $\cb{h_{11}g_{1,i},h_{12}g_{2,n_d}}_{i=1}^{n_d}$ are \textit{rationally independent}\footnote{A set of $n$ complex numbers $\{c_1,\ldots,c_n\}\subset\mathbb C^n$ is said to be rationally independent if the only solution to the equation $\sum_{i=1}^{n}p_ic_i=0$ with integer coefficients $\{p_1,\ldots,p_n\}\in\mathbb Z^n$ is the trivial solution $p_i=0$ for all $i\in[n]$.}, and analogously for  $\cb{h_{22}g_{2,i},h_{21}g_{1,n_d}}_{i=1}^{n_d}$. Therefore, for each layer, by \cite[Theorem 3]{motahari2014real}, there exists a code, of rate
\begin{IEEEeqnarray}{rCl}
	R=\rb{1-e^{-\frac{d_{\text{min}}}{8}}}\log Q-1
\end{IEEEeqnarray}
such that UE 1 can decode layers $\cb{a_1,a_2+b_1,\ldots,a_{n_d}+b_{n_d-1},b_{n_d}}$, while UE 2 decodes  $\{b_1,b_2+a_1,\ldots,b_{n_d}+a_{n_d-1},a_{n_d}\}$. Furthermore, it follows from \cite[Theorem 4]{maddah2010degrees} that, for almost all CSI $\mathbf{H}$, the minimum distance $d_\text{min}$ satisfies the inequality
\begin{IEEEeqnarray}{rCl}
	d_\text{min}&>&\dfrac{A}{(2Q)^{\frac{n_d-1}{2}+\epsilon}},
\end{IEEEeqnarray}
for every $\epsilon>0$. 

Next, we choose parameters $A$ and $Q$ such that, on the one hand, the average power constraint is satisfied, and, on the other hand, the minimum distance $d_{\text{min}}$ grows with $P$. Let $A=Q^{\frac{n_d-1}{2}+\epsilon'}$ and $Q=\rho(\mathbf{H},n_d)P^{\frac{1}{n_d+1+2\epsilon'}}$, with $\epsilon'>\epsilon$ and with a constant $0<\rho(\mathbf{H},n_d)<1$ that depends only on the CSI and the number of layers. Note that, since $Q$ should be an integer, additional rounding is required, but this does not affect the high-SNR analysis and is therefore omitted. With these choices, we have
\begin{IEEEeqnarray}{rCl}
	d_\text{min}&>&\frac{\rho(\mathbf{H},n_d)^{\epsilon'-\epsilon}P^{(\epsilon'-\epsilon)/(n_d+1+2\epsilon')}}{2^{(n_d-1)/2+\epsilon}}\xrightarrow[P\rightarrow\infty]{}\infty,
\end{IEEEeqnarray}
and
\begin{IEEEeqnarray}{rCl}
	(AQ)^2=\rb{Q^{\frac{n_d+1}{2}+\epsilon'}}^2=\rho(\mathbf{H},n_d)^{n_d+1+2\epsilon'}P,
\end{IEEEeqnarray}
and hence the role of $\rho(\mathbf{H},n_d)$ is to maintain the average power constraint by accounting for the precoder gains and the number of layers. It follows that, for high SNR, we have the limit
\begin{IEEEeqnarray}{rCl}
	\lim_{P\rightarrow\infty}\frac{R}{\log P}=\lim_{P\rightarrow\infty}\frac{\log Q}{\log P}=\frac{1}{n_d+1+2\epsilon'}.
\end{IEEEeqnarray}

As was described in Sec. \ref{sec:real_IA}, UE 1 transmits the message $v_1=\{a_2+b_1,a_4+b_3,\ldots,a_{n_d-1}+b_{n_d-2}\}$ to UE 2, while UE 2 transmits $v_2=\{b_2+a_1,b_4+a_3,\ldots,b_{n_d-1}+a_{n_d-2}\}$ to UE 1, over the D2D links. These include $(n_d-1)/2$ layers, each taking one of $2Q$ values. As a result, UE 1 is able to decode $\{a_1,b_2,a_3,b_4,\ldots,a_{n_d},b_{n_d}\}$ while UE 2 decodes $\{b_1,a_2,b_3,a_4,\ldots,b_{n_d},a_{n_d}\}$, at the expense of the D2D latency 
\begin{IEEEeqnarray}{rCl}
	T_D=T_E\cdot\frac{\log(2Q)\cdot(n_d-1)/2}{r_D\log P}.
\end{IEEEeqnarray}
Since each UE decodes $(n_d+1)/2$ layers from one EN and $(n_d-1)/2$ from the other, the number of channels uses required to satisfy the UEs' demands is
\begin{IEEEeqnarray}{rCl}
	T_E=\frac{L/2}{R\cdot \frac{n_d-1}{2}}.
\end{IEEEeqnarray}
Thus, the resulting NDT is
\begin{IEEEeqnarray}{rCl}
	\delta(\mu=1/2,r_F,r_D)&=&\lim_{P\rightarrow\infty}\lim_{L\rightarrow\infty}\frac{(T_E+T_D)\log P}{L}\IEEEnonumber\\
	&=&\lim_{P\rightarrow\infty}\frac{\log P}{(n_d-1)R}\sqb{1+\frac{n_d-1}{2r_D}\cdot\frac{\log(2Q)}{\log P}}\IEEEnonumber\\
	&=&\frac{n_d+1+2\epsilon'}{n_d-1}\sqb{1+\frac{1}{2r_D}\cdot\frac{n_d-1}{n_d+1+2\epsilon'}},
\end{IEEEeqnarray}
and for sufficiently small $\epsilon'$ and sufficiently large $n_d$, we can get arbitrarily close to $\delta_X$ \eqref{eq:x_ch_ub}.

\section{Lower bound on the minimum MDT}\label{sec:appendix_lb}
Here we prove the converse result that demonstrates the optimality of the proposed scheme. To this end, without loss of generality, assume that UE 1 and UE 2 request files $f_1$ and $f_2$, respectively. Define $\mathbf{f}_n^N\triangleq \rb{f_n,f_{n+1},\ldots,f_N}$. The lower bounds are obtained by identifying subsets of information resources from which, for high-SNR, both files should be reliably decoded when a feasible policy is implemented. As discussed, we specifically consider in turn the subsets $\cb{s_1,s_2,\mathbf{u}_{1},\mathbf{u}_{2}}$, $\cb{\mathbf{y}_1,\mathbf{y}_2}$ and $\cb{s_1,\mathbf{u}_{1},\mathbf{y}_1,\mathbf{v}_2}$.

Let $\epsilon_L$ be a function of $L$, independent of $P$, such that $\epsilon_L\rightarrow 0$ as $L\rightarrow\infty$. Since the D2D messages $\mathbf{v}_1$ and $\mathbf{v}_2$ are functions of the downlink received signals $\mathbf{y}_1$ and $\mathbf{y}_2$, then we have
\begin{IEEEeqnarray}{rCl}
	H\AgivenB{f_1,f_2}{\mathbf{y}_1,\mathbf{y}_2,\mathbf{f}_3^N}&=&H\AgivenB{f_1,f_2}{\mathbf{y}_1,\mathbf{y}_2,\mathbf{v}_1,\mathbf{v}_2,\mathbf{f}_3^N}\IEEEnonumber\\
	&\leq&H\AgivenB{f_1}{\mathbf{y}_1,\mathbf{v}_2}+H\AgivenB{f_2}{\mathbf{y}_2,\mathbf{v}_1}\IEEEnonumber\\
	&\leq& 2\epsilon_L,
\end{IEEEeqnarray}
where the last step follows from Fano's inequality since each file $f_k$ can be decoded from $\{\mathbf{y}_k,\mathbf{v}_{k'}\}$, $k,k'\in[2]$, $k'\neq k$, by the definition of feasible policies. Therefore, when considering subsets $\cb{s_1,s_2,\mathbf{u}_{1},\mathbf{u}_{2}}$ and $\cb{\mathbf{y}_1,\mathbf{y}_2}$, we can readily use the derivations of \cite{tandon2016cloud} to get \textit{Inequality 2} and \textit{Inequality 3} of \cite[Appendix A]{tandon2016cloud}, that is,
\begin{IEEEeqnarray}{rCl}
	T_F\log(P)\geq \frac{(1-2\mu)L}{r_F}-\frac{L\epsilon_L}{r_F},
\end{IEEEeqnarray}
and 
\begin{IEEEeqnarray}{rCl}
	T_E\log(P)\geq L-L\epsilon_L.
\end{IEEEeqnarray}

For the subset $\cb{s_1,\mathbf{u}_{1},\mathbf{y}_1,\mathbf{v}_2}$, we consider the following equality:
\begin{IEEEeqnarray}{rCl}\label{eq:subset3}
	2L&=&H\AgivenB{f_1,f_2}{\mathbf{f}_3^N}\IEEEnonumber\\
	&=&I\AgivenB{f_1,f_2;s_1,\mathbf{u}_1,\mathbf{y}_1,\mathbf{v}_2}{\mathbf{f}_3^N}+H\AgivenB{f_1,f_2}{s_1,\mathbf{u}_1,\mathbf{y}_1,\mathbf{v}_2,\mathbf{f}_3^N}.
\end{IEEEeqnarray}
The first term in \eqref{eq:subset3} can be bounded as
\begin{IEEEeqnarray}{rCl}
	I\AgivenB{f_1,f_2;s_1,\mathbf{u}_1,\mathbf{y}_1,\mathbf{v}_2}{\mathbf{f}_3^N}
	&=&I\AgivenB{f_1,f_2;\mathbf{y}_1}{\mathbf{f}_3^N}+I\AgivenB{f_1,f_2;\mathbf{v}_2}{\mathbf{y}_1,\mathbf{f}_3^N}\IEEEyesnumber\IEEEyessubnumber\label{eq:chain_rule}\\
	&&+I\AgivenB{f_1,f_2;s_1,\mathbf{u}_1}{\mathbf{y}_1,\mathbf{v}_2,\mathbf{f}_3^N}\IEEEnonumber\\
	&\leq&T_E\log\rb{1+\lambda P}+I\AgivenB{f_1,f_2;\mathbf{v}_2}{\mathbf{y}_1,\mathbf{f}_3^N}\IEEEyessubnumber\label{eq:lemma5}\\
	&&+I\AgivenB{f_1,f_2;s_1,\mathbf{u}_1}{\mathbf{y}_1,\mathbf{v}_2,\mathbf{f}_3^N}\IEEEnonumber\\
	&\leq&T_E\log\rb{1+\lambda P}+I\AgivenB{f_1,f_2;\mathbf{v}_2}{\mathbf{y}_1,\mathbf{f}_3^N}\IEEEnonumber\\
	&&+I\AgivenB{f_1,f_2;f_1,s_1,\mathbf{u}_1}{\mathbf{y}_1,\mathbf{v}_2,\mathbf{f}_3^N}\IEEEnonumber\\
	&=&T_E\log\rb{1+\lambda P}+I\AgivenB{f_1,f_2;\mathbf{v}_2}{\mathbf{y}_1,\mathbf{f}_3^N}\IEEEnonumber\\
	&&+I\AgivenB{f_1,f_2;f_1}{\mathbf{y}_1,\mathbf{v}_2,\mathbf{f}_3^N}+I\AgivenB{f_1,f_2;s_1,\mathbf{u}_1}{\mathbf{y}_1,\mathbf{v}_2,f_1,\mathbf{f}_3^N}\IEEEnonumber\\
	&\leq&T_E\log\rb{1+\lambda P}+L\epsilon_L\IEEEnonumber\\
	&&+I\AgivenB{f_1,f_2;\mathbf{v}_2}{\mathbf{y}_1,\mathbf{f}_3^N}+I\AgivenB{f_1,f_2;s_1,\mathbf{u}_1}{\mathbf{y}_1,\mathbf{v}_2,f_1,\mathbf{f}_3^N}\IEEEyessubnumber\label{eq:fano3},
\end{IEEEeqnarray}
where \eqref{eq:chain_rule} follows from the chain rule for mutual information; inequality \eqref{eq:lemma5} follows from \cite[Lemma 5]{sengupta2016fog} with $\lambda\triangleq\max_{k\in[2]}|h_{k1}+h_{k2}|^2$; and inequality \eqref{eq:fano3} follows from Fano's inequality since $I\AgivenB{f_1,f_2;f_1}{\mathbf{y}_1,\mathbf{v}_2,\mathbf{f}_3^N}\leq H\AgivenB{f_1}{\mathbf{y}_1,\mathbf{v}_2}$.
The mutual information $I\AgivenB{f_1,f_2;s_1,\mathbf{u}_1}{\mathbf{y}_1,\mathbf{v}_2,f_1,\mathbf{f}_3^N}$ in \eqref{eq:fano3} can be bounded as
\begin{IEEEeqnarray}{rCl}
	I\AgivenB{f_1,f_2;s_1,\mathbf{u}_1}{\mathbf{y}_1,\mathbf{v}_2,f_1,\mathbf{f}_3^N}&\leq&H\AgivenB{s_1,\mathbf{u}_1}{\mathbf{y}_1,\mathbf{v}_2,f_1,\mathbf{f}_3^N}\IEEEnonumber\\
	&\leq&H\AgivenB{s_1,\mathbf{u}_1}{f_1,\mathbf{f}_3^N}\IEEEnonumber\\
	&=&H\AgivenB{s_{1,1},\ldots,s_{1,N},\mathbf{u}_1}{f_1,\mathbf{f}_3^N}\IEEEyesnumber\IEEEyessubnumber\label{eq:eq_def_s1}\\
	&=&H\AgivenB{s_{1,2},\mathbf{u}_1}{f_1,\mathbf{f}_3^N}\IEEEyessubnumber\label{eq:eq_cache}\\
	&\leq&H\rb{s_{1,2},\mathbf{u}_1}\IEEEnonumber\\
	&\leq&H\rb{s_{1,2}}+H\rb{\mathbf{u}_1}\IEEEnonumber\\
	&\leq&\mu L+r_FT_F\log(P)\IEEEnonumber,
\end{IEEEeqnarray}
where \eqref{eq:eq_def_s1} follows from the definition of $s_1$; and equality \eqref{eq:eq_cache} follows from the fact that $s_{1,n}$ is a function of $f_n$ \eqref{eq:caching_policy}.
Finally, the term $I\AgivenB{f_1,f_2;\mathbf{v}_2}{\mathbf{y}_1,\mathbf{f}_3^N}$ in \eqref{eq:fano3} can be bounded as
\begin{IEEEeqnarray}{rCl}
	I\AgivenB{f_1,f_2;\mathbf{v}_2}{\mathbf{y}_1,\mathbf{f}_3^N}&=&H\AgivenB{\mathbf{v}_2}{\mathbf{y}_1,\mathbf{f}_3^N}-H\AgivenB{\mathbf{v}_2}{\mathbf{y}_1,\mathbf{f}_1^N}\IEEEnonumber\\
	&\leq&H\AgivenB{\mathbf{v}_2}{\mathbf{y}_1,\mathbf{f}_3^N}\IEEEnonumber\\
	&\leq&H\rb{\mathbf{v}_2}\IEEEnonumber\\
	&\leq&r_DT_D\log(P).\label{eq:bound_v_given_y}
\end{IEEEeqnarray}
Therefore, the first term in \eqref{eq:subset3} is bounded as
\begin{IEEEeqnarray}{rCl}\label{eq:I_subset3}
	I\AgivenB{f_1,f_2;s_1,\mathbf{u}_1,\mathbf{y}_1,\mathbf{v}_2}{\mathbf{f}_3^N}
	&\leq&T_E\log\rb{1+\lambda P}+L\epsilon_L+r_DT_D\log(P)+\mu L+r_FT_F\log(P).
\end{IEEEeqnarray}

Next, the second term in \eqref{eq:subset3} can be bounded as
\begin{IEEEeqnarray}{rCl}\label{eq:bound_s_u_v}
	H\AgivenB{f_1,f_2}{s_1,\mathbf{u}_1,\mathbf{y}_1,\mathbf{v}_2,\mathbf{f}_3^N}
	&=&H\AgivenB{f_1,f_2}{s_1,\mathbf{u}_1,\mathbf{y}_1,\mathbf{y}_2,\mathbf{v}_2,\mathbf{f}_3^N}+I\AgivenB{f_1,f_2;\mathbf{y}_2}{s_1,\mathbf{u}_1,\mathbf{y}_1,\mathbf{v}_2,\mathbf{f}_3^N}\IEEEnonumber\\
	&\leq&H\AgivenB{f_1,f_2}{\mathbf{y}_1,\mathbf{y}_2}+I\AgivenB{f_1,f_2;\mathbf{y}_2}{s_1,\mathbf{u}_1,\mathbf{y}_1,\mathbf{v}_2,\mathbf{f}_3^N}\IEEEnonumber\\
	&\leq&L\epsilon_L+I\AgivenB{f_1,f_2;\mathbf{y}_2}{s_1,\mathbf{u}_1,\mathbf{y}_1,\mathbf{v}_2,\mathbf{f}_3^N}\IEEEyesnumber\IEEEyessubnumber\label{eq:fano2}\\
	&=&L\epsilon_L+h\AgivenB{\mathbf{y}_2}{s_1,\mathbf{u}_1,\mathbf{y}_1,\mathbf{v}_2,\mathbf{f}_3^N}-h\AgivenB{\mathbf{y}_2}{s_1,\mathbf{u}_1,\mathbf{y}_1,\mathbf{v}_2,\mathbf{f}_1^N}\IEEEnonumber\\
	&=&L\epsilon_L+h\AgivenB{\mathbf{y}_2}{s_1,\mathbf{u}_1,\mathbf{x}_1,\mathbf{y}_1,\mathbf{v}_2,\mathbf{f}_3^N}-h\rb{\mathbf{z}_2}\IEEEyessubnumber\label{eq:x_func_u_s}\\
	&\leq& L\epsilon_L+h\AgivenB{\mathbf{y}_2}{\mathbf{x}_1,\mathbf{y}_1}-h\rb{\mathbf{z}_2},\IEEEyessubnumber\label{eq:cond_red_ent}
\end{IEEEeqnarray}
where \eqref{eq:fano2} follows from Fano's inequality; equality \eqref{eq:x_func_u_s} follows because the transmitted signal $\mathbf{x}_1$ is a function of $s_1$ and $\mathbf{u}_1$ as well as from the fact that given all the files $\mathbf{f}_1^N$, the channel outputs $\mathbf{y}_2$ are a function of the channel noises $\mathbf{z}_2$ only; and inequality \eqref{eq:cond_red_ent} follows the conditioning reduces entropy property.
Since  the channel matrix $\mathbf{H}$ is drawn from a continuous distribution, then almost surely the received signals can be related as
\begin{IEEEeqnarray}{rCl}
	\mathbf{y}_2&=&\frac{h_{22}}{h_{12}}\mathbf{y}_1-\rb{h_{21}-\frac{h_{11}h_{22}}{h_{12}}}\mathbf{x}_1-\frac{h_{22}}{h_{12}}\mathbf{z}_1+\mathbf{z}_2,
\end{IEEEeqnarray}
and hence $\mathbf{y}_2+\tilde{\mathbf{z}}_2$, where $\tilde{\mathbf{z}}_2\triangleq (h_{22}/h_{12})\mathbf{z}_1-\mathbf{z}_2$, is almost surely given by 
\begin{IEEEeqnarray}{rCl}
	\mathbf{y}_2+\tilde{\mathbf{z}}_2&=&\frac{h_{22}}{h_{12}}\mathbf{y}_1-\rb{h_{21}-\frac{h_{11}h_{22}}{h_{12}}}\mathbf{x}_1,
\end{IEEEeqnarray}
that is, the signal $\mathbf{y}_2+\tilde{\mathbf{z}}_2$ is a function of $\mathbf{y}_1$ and $\mathbf{x}_1$. Therefore, continuing from \eqref{eq:cond_red_ent}, we have
\begin{IEEEeqnarray}{rCl}\label{eq:H_subset3}
	H\AgivenB{f_1,f_2}{s_1,\mathbf{u}_1,\mathbf{y}_1,\mathbf{v}_2,\mathbf{f}_3^N}&\leq&L\epsilon_L+h\AgivenB{\mathbf{y}_2}{\mathbf{x}_1,\mathbf{y}_1}-h\rb{\mathbf{z}_2}\IEEEnonumber\\
	&=&L\epsilon_L+h\AgivenB{\mathbf{y}_2}{\mathbf{x}_1,\mathbf{y}_1,\mathbf{y}_2+\tilde{\mathbf{z}}_2}-h\rb{\mathbf{z}_2}\IEEEnonumber\\
	&\leq&L\epsilon_L+h\AgivenB{\mathbf{y}_2}{\mathbf{y}_2+\tilde{\mathbf{z}}_2}-h\rb{\mathbf{z}_2}\IEEEnonumber\\
	&=&L\epsilon_L+h\AgivenB{\mathbf{y}_2-(\mathbf{y}_2+\tilde{\mathbf{z}}_2)}{\mathbf{y}_2+\tilde{\mathbf{z}}_2}-h\rb{\mathbf{z}_2}\IEEEnonumber\\
	&=&L\epsilon_L+h\AgivenB{\tilde{\mathbf{z}}_2}{\mathbf{y}_2+\tilde{\mathbf{z}}_2}-h\rb{\mathbf{z}_2}\IEEEnonumber\\
	&\leq&L\epsilon_L+h\rb{\tilde{\mathbf{z}}_2}-h\rb{\mathbf{z}_2}\IEEEnonumber\\
	&=&L\epsilon_L+T_E\log\rb{1+\frac{|h_{22}|^2}{|h_{12}|^2}}.
\end{IEEEeqnarray}
Substituting \eqref{eq:I_subset3} and \eqref{eq:H_subset3} in \eqref{eq:subset3} leads to the inequality
\begin{IEEEeqnarray}{rCl}
	2L&\leq&T_E\log\rb{1+\lambda P}+L\epsilon_L+r_DT_D\log(P)+\mu L+r_FT_F\log(P)+L\epsilon_L+T_E\log\rb{1+\frac{|h_{22}|^2}{|h_{12}|^2}}.
\end{IEEEeqnarray}

To summarize, we have the following three inequalities:
\begin{IEEEeqnarray}{rCl}
	\rb{T_E+r_FT_F+r_DT_D}\log(P)&\geq&\rb{2-\mu}L-2L\epsilon_L-T_E\log\rb{\lambda+\frac{1}{P}}-T_E\log\rb{1+\frac{|h_{22}|^2}{|h_{12}|^2}},\label{eq:lb_inequality1}\\
	T_F\log(P)&\geq&\frac{L\rb{1-2\mu}}{r_F}-\frac{L\epsilon_L}{r_F},\label{eq:lb_inequality2}\\
	T_E\log(P)&\geq& L-L\epsilon_L\label{eq:lb_inequality3}.
\end{IEEEeqnarray}
For $0\leq r_F,r_D\leq 1$, we use \eqref{eq:lb_inequality1} to get the lower bound
\begin{IEEEeqnarray}{rCl}
	\delta^*(\mu,r_F,r_D)&=&\lim_{P\rightarrow\infty}\lim_{L\rightarrow\infty}\frac{\rb{T_E+T_F+T_D}\log(P)}{L}\IEEEnonumber\\
	&\geq&\lim_{P\rightarrow\infty}\lim_{L\rightarrow\infty}\frac{\rb{T_E+r_FT_F+r_DT_D}\log(P)}{L}\IEEEnonumber\\
	&\geq&2-\mu.
\end{IEEEeqnarray}
Furthermore, by multiplying \eqref{eq:lb_inequality2} by $(1-r_F)$ and adding the resulting bound to \eqref{eq:lb_inequality1}, we get the lower bound
\begin{IEEEeqnarray}{rCl}
	\delta^*(\mu,r_F,r_D)&=&\lim_{P\rightarrow\infty}\lim_{L\rightarrow\infty}\frac{\rb{T_E+T_F+T_D}\log(P)}{L}\IEEEnonumber\\
	&\geq&\lim_{P\rightarrow\infty}\lim_{L\rightarrow\infty}\frac{\rb{T_E+T_F+r_DT_D}\log(P)}{L}\IEEEnonumber\\
	&=&	\lim_{P\rightarrow\infty}\lim_{L\rightarrow\infty}\frac{\rb{T_E+r_FT_F+r_DT_D+\rb{1-r_F}T_F}\log(P)}{L}\IEEEnonumber\\
	&\geq&2-\mu+\frac{1-r_F}{r_F}\rb{1-2\mu}\IEEEnonumber\\
	&=&1+\mu+\frac{1-2\mu}{r_F}.
\end{IEEEeqnarray}
Thus, for $0\leq r_F,r_D\leq 1$, we have
\begin{IEEEeqnarray}{rCl}
	\delta^*(\mu,r_F,r_D)&\geq&\max\cb{1+\mu+\frac{1-2\mu}{r_F},2-\mu}\label{eq:rD<1_lb1}.
\end{IEEEeqnarray}

For $r_F>\max\cb{1,r_D}$, i.e., $r_F>1$ and $0\leq r_D\leq r_F$, by multiplying \eqref{eq:lb_inequality3} by $\rb{r_F-1}$ and adding the resulting bound to \eqref{eq:lb_inequality1}, we get
\begin{IEEEeqnarray}{rCl}
	\rb{r_FT_E+r_FT_F+r_DT_D}\log(P)&\geq& \rb{1-\mu+r_F}L-\rb{1+r_F}L\epsilon_L\IEEEnonumber\\
	&&-T_E\log\rb{\lambda+\frac{1}{P}}-T_E\log\rb{\frac{|h_{22}|^2}{|h_{12}|^2}}.\label{eq:lb_calc1}
\end{IEEEeqnarray}
Moreover, for $r_D<r_F$, we have
\begin{IEEEeqnarray}{rCl}
	r_F\rb{T_E+T_F+T_D}\log(P)&\geq&\rb{r_FT_E+r_FT_F+r_DT_D}\log(P),
\end{IEEEeqnarray}
which, together with \eqref{eq:lb_calc1}, results in the lower bound
\begin{IEEEeqnarray}{rCl}
	\delta^*(\mu,r_F,r_D)&\geq&1+\frac{1-\mu}{r_F}\label{eq:rD<1_lb2}.
\end{IEEEeqnarray}

Next, for $r_D>\max\cb{1,r_F}$, i.e., $r_D>1$ and $0\leq r_F< r_D$, similar to \eqref{eq:lb_calc1}-\eqref{eq:rD<1_lb2}, by multiplying \eqref{eq:lb_inequality3} by $\rb{r_D-1}$ and adding the resulting bound to \eqref{eq:lb_inequality1}, we get the lower bound
\begin{IEEEeqnarray}{rCl}
	\delta^*(\mu,r_F,r_D)&\geq&1+\frac{1-\mu}{r_D}.
\end{IEEEeqnarray}
Moreover, we use $\eqref{eq:lb_inequality1} + \rb{r_D-r_F} \times \eqref{eq:lb_inequality2} + \rb{r_D-1} \times \eqref{eq:lb_inequality3}$ to get the inequality
\begin{IEEEeqnarray}{rCl}
	r_D\rb{T_E+T_F+T_D}\log(P)&\geq&L\sqb{1-\mu+r_D+\frac{r_D-r_F}{r_F}\rb{1-2\mu}}\IEEEnonumber\\
	&&-L\epsilon_L\sqb{1+r_D+\frac{r_D-r_F}{2r_F}}\IEEEnonumber\\
	&&-T_E\log\rb{\lambda+\frac{1}{P}}r_D-T_E\log\rb{1+\frac{|h_{22}|^2}{|h_{12}|^2}}.
\end{IEEEeqnarray}
Hence, we have an additional lower bound
\begin{IEEEeqnarray}{rCl}
	\delta^*(\mu,r_F,r_D)&\geq&1+\frac{1-\mu}{r_D}+\frac{r_D-r_F}{r_Dr_F}\rb{1-2\mu}\IEEEnonumber\\
	&=&1+\frac{\mu}{r_D}+\frac{1-2\mu}{r_F}.
\end{IEEEeqnarray}
Overall, for $r_D>\max\cb{1,r_F}$, the lower bound is given by
\begin{IEEEeqnarray}{rCl}
	\delta^*(\mu,r_F,r_D)&\geq&\max\cb{1+\frac{\mu}{r_D}+\frac{1-2\mu}{r_F},1+\frac{1-\mu}{r_D}},
\end{IEEEeqnarray}
which completes the proof.

% Bibiliography
\bibliographystyle{IEEEtran}
\bibliography{IEEEabrv,bib/myBib}

\end{document}